\documentclass[twocolumn,american,aps,superscriptaddress,floatfix,final,showpacs,prb,reprint]{revtex4-1}
\usepackage{mathptmx}

\usepackage[T1]{fontenc}
\usepackage[latin9]{inputenc}
\usepackage{color}
\usepackage{float}
\usepackage{textcomp}
\usepackage{amstext}
\usepackage{amssymb}
\usepackage{graphicx}
\usepackage{esint}

\makeatletter

\@ifundefined{textcolor}{}
{%
  \definecolor{BLACK}{gray}{0}
  \definecolor{WHITE}{gray}{1}
  \definecolor{RED}{rgb}{1,0,0}
  \definecolor{GREEN}{rgb}{0,1,0}
  \definecolor{BLUE}{rgb}{0,0,1}
  \definecolor{CYAN}{cmyk}{1,0,0,0}
  \definecolor{MAGENTA}{cmyk}{0,1,0,0}
  \definecolor{YELLOW}{cmyk}{0,0,1,0}
}
 
\renewcommand\[{\begin{equation}}
\renewcommand\]{\end{equation}}
\usepackage{bm}

\@ifundefined{showcaptionsetup}{}{%
 \PassOptionsToPackage{caption=false}{subfig}}
\usepackage{subfig}
\makeatother

\usepackage{babel}
\begin{document}
\title{Unusually thick metal-insulator domain walls around the Mott point}
\author{M. Y. Su\'arez-Villagr\'an}
\affiliation{Department of Physics and Texas Center for Superconductivity, University
of Houston, Houston, TX 77204-5005 USA}
\author{N. Mitsakos}
\affiliation{Department of Mathematics, University of Houston, Houston, TX 77204-5008
USA}
\author{Tsung-Han Lee}
\affiliation{Department of Physics and National High Magnetic Field Laboratory,
Florida State University, Tallahassee, Florida 32306, USA}
\affiliation{Physics and Astronomy Department, Rutgers University, Piscataway,
New Jersey 08854, USA}
\author{J. H. Miller, Jr.}
\affiliation{Department of Physics and Texas Center for Superconductivity, University
of Houston, Houston, TX 77204-5005 USA}
\author{E. Miranda}
\affiliation{Gleb Wataghin Physics Institute, The University of Campinas, Rua
S\'ergio Buarque de Holanda, 777, CEP 13083-859, Campinas, Brazil}
\author{V. Dobrosavljevi\'{c}}
\affiliation{Department of Physics and National High Magnetic Field Laboratory,
Florida State University, Tallahassee, Florida 32306, USA}

\begin{abstract}
Many Mott systems feature a first-order metal-insulator transition at finite
temperatures, with an associated phase coexistence region displaying
inhomogeneities and local phase separation. Here one typically finds ``bubbles'' 
or domains of the respective phases, which are separated by surprisingly thick, or fat,
domain walls, as revealed both by imaging experiments and recent theoretical modeling. 
To gain insight into this unexpected behavior, we perform a systematic  model
study of the structure of such metal-insulator domain walls around the Mott point, within
the Dynamical Mean-Field Theory framework. Our study reveals that a mechanism producing 
such ``fat'' domain walls can be traced to strong magnetic frustration, which is 
expected to be a robust feature of ``spin-liquid'' Mott systems.
\end{abstract}
\maketitle

\section{\textit{\emph{Introduction}}}

\label{sec:intro}

The  Mott (interaction-driven) metal-insulator transition represents
one of the most important phenomena in strongly correlated electron
systems.\cite{Mott1990} It was first recognized in a number of transition-metal
oxides,\cite{Lederer,McWhan,Mazzaferro,Limelette03,M.Qazilbash} and has been 
brought to notoriety with the discovery of 
the cuprate high-$T_{c}$ superconductors,\cite{lee2004doping,cai} which 
raised much controversy surrounding its character and the underlying physical 
processes. One popular viewpoint, going back to early ideas of Slater, assumes that the 
key mechanism producing the Mott insulating state follows from magnetic order
simply rearranging the band structure. An alternative perspective, pioneered by the seminal 
ideas of Mott and Anderson, argues that strong Coulomb repulsion may arrest the electronic motion 
even in the absence of magnetic order. Several theoretical scenarios 
\cite{Hubbard1957,Hubbard1958,Brinkman1970, A.Georges1996} 
have been proposed for the vicinity 
of the Mott point, but for many years the controversy remained unresolved. 

More recent experiments have demonstrated \cite{kanoda2005prl} that a sharp Mott transition is indeed
possible even in the absence of any magnetic order. Physically, this possibility is realized  in systems 
where sufficiently strong magnetic frustration\cite{Powell2011RoP} can suppress magnetic order down  to low enough temperatures, thus 
revealing the ``pure'' Mott point. Precisely such behavior is found in a class $\kappa$-organic 
``spin-liquid'' materials,\cite{dressel2020advphys} which have been recently  recognized\cite{vollhardt2019jpspc} 
as the ideal realization of a 
single-band Hubbard model on a triangular lattice, allowing a remarkably detailed insight into the Mott transition region. While the  intermediate-temperature metal-insulator crossover revealed\cite{Furukawa2015} some striking 
aspects of quantum  criticality\cite{H.Terletska} around the quantum Widom line, \cite{Vucicevic2013,Pustogow2018} the transition was found to assume weakly first-order character at the lowest temperatures. Spectacular anomalies in dielectric response were observed\cite{pustogow2021low} within the associated phase coexistence region, revealing percolative phase separation. 

Remarkably, most qualitative and even semi-quantitative aspects of the behavior\cite{dressel2020advphys}  observed around the Mott point were found to validate the predictions of Dynamical Mean-Field Theory (DMFT). \cite{Pruschke1995,A.Georges1996} This theoretical approach,
which gained considerable popularity in recent years,\cite{Kotliar2004} 
focuses on the local effects of strong Coulomb repulsion, while disregarding certain nonlocal
processes associated with inter-site magnetic correlations and/or fluctuating magnetic orders. 
It reconciled the earlier theories of Hubbard\cite{Hubbard1957,Hubbard1958}
with the viewpoint of Brinkman and Rice,\cite{Brinkman1970} leading to a consistent non-perturbative 
picture of the Mott point. 

\begin{figure}[H]
\centering{}\includegraphics[scale=1]{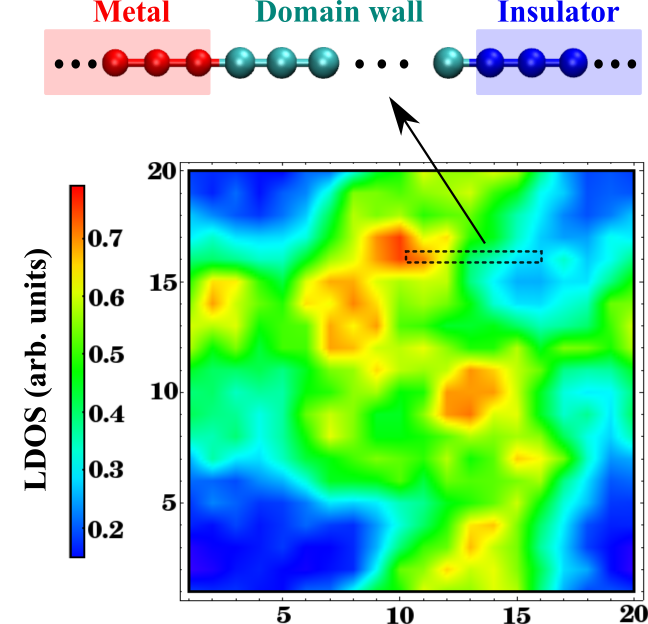}\caption{\label{bubbles} Spatial fluctuations of the 
local density of states (LDOS) found in a recent simulation of a moderately disordered 
two-dimensional Mott system, within 
the metal-insulator phase coexistence region.\cite{martha} Note how the domain walls (green)
covers a substantial area of the image, separating metallic (red) from the Mott-insulating 
domains (blue). Similar patterns have also been found in earlier imaging experiments on certain Mott oxides.\cite{M.Qazilbash}}
\end{figure}

While many aspects of crystalline Mott materials prove to be successfully described and interpreted from the perspective of DMFT, the situation is more
complicated in the presence of disorder,\cite{Miranda2005,Tanaskovi'c2003a,Dobrosavljevi'c2003,M.C.O.Aguiar2004,Aguiar2005,aguiaretal09,Andrade2009a,andradeetal10} which breaks translational invariance. 
These effects are most pronounced, but least understood, within the metal-insulator phase coexistence 
region. Here even moderate disorder creates nucleation centers for the respective phases,
leading to nano-scale phase separation. Some aspects of this behavior proved 
possible to be described from the perspective of a phenomenological percolation picture, 
including the effects of inhomogeneities caused by thermal fluctuations around the critical end-point,\cite{PhysRevLett.100.026408} as well as the colossal dielectric response at lower
temperatures.\cite{pustogow2021low} 

A closer look at the microscopic structure of the corresponding patterns, however, revealed 
various puzzling features. Hints of remarkably complex behavior were provided by very recent
large-scale numerical modeling\cite{martha} of disordered Mott systems, as well as experimentally by 
nano-scale imaging of some Mott materials.\cite{M.Qazilbash} A typical
situation is illustrated in Fig.~\ref{bubbles}, where we reproduce a result of our recent theoretical
study\cite{martha} of this regime. Here we see clearly defined metallic domains separated from 
Mott-insulating regions by surprisingly thick domain walls, which in some cases cover a large 
fraction of the system volume (area). Remarkably, very similar fat domain walls were also observed 
in certain experiments,\cite{M.Qazilbash} suggesting robust new physics. This finding immediately brings 
into question the conventional percolation picture, where the domain walls are assumed to play only a secondary role. This observation also brings forth several important physical questions, which are 
the primary motivation for this work: (1) What is the physical reason for having rather thick or fat 
domain walls, and under which conditions is this expected to hold? (2) What are the physical 
properties of such ``domain wall matter'', and how should they affect the physical observables? 

In this study we present a detailed theoretical investigation of the structure of such domain 
walls not only in the vicinity to the critical end-point\cite{LeeHou} (where one generally expects them to be thick), but also across the entire phase-coexistence region. Our results establish that, under certain conditions, such domain walls can remain thick in the entire range of temperatures, and reveal the underlying mechanism, at least within the DMFT picture we adopt. We argue that strong magnetic frustration acts as a 
key physical ingredient affecting the properties of such domain walls, also suggesting ways to 
further control their properties in ``materials by design''.\cite{Adler_2018}

\section{\textit{\emph{Model and method}}}

\label{sec:method}

To microscopically investigate metal-insulator domain walls in the vicinity of the Mott point, we focus on a single-band half-filled Hubbard model, as given by the Hamiltonian

\begin{equation}
H=-t\sum_{i\sigma}\left[c_{i\sigma}^{\dagger}c_{\left(i+1\right)\sigma}^{\phantom{\dagger}}+\mathrm{h.c.}\right]+U\sum_{i}\left(n_{i\uparrow}-\frac{1}{2}\right)\left(n_{i\downarrow}-\frac{1}{2}\right),\label{eq:hamiltonian}
\end{equation}
where $c_{i\sigma}^{\dagger}\left(c_{i\sigma}\right)$ is the creation
(annihilation) operator of an electron with spin projection $\sigma$
on site $i$, $t$ is the hopping amplitude between nearest neighbors,
$U$ is the on-site Coulomb repulsion, and $n_{i\sigma}=c_{i\sigma}^{\dagger}c_{i\sigma}$.
We work in units such that $\hslash=k_{B}=a=1$, where $a$ is the lattice
spacing. Energy will generally be given in the units of half-band width $D$, which for our half-filled situation is also the Fermi energy.

In general terms, a domain wall in  $d$ dimensions is a ($d-1$)-dimensional
surface separating two thermodynamic phases. To examine its basic properties, 
we follow a standard procedure\cite{Nigel} in assuming it to be flat, i.e. that its
spatial variation across the wall is the only relevant one. To further simplify the calculation,
we  take advantage of the the well-established fact that,  within the DMFT formulation we employ, the detailed form of the electronic band structure does not qualitatively affect the results.\cite{A.Georges1996,Helmes2008,potthoff} This allows us follow the same strategy as in standard theories for domain walls, and reduce the problem to solving a one-dimensional model with appropriate boundary conditions on each end representing the respective thermodynamic phases. 

\noindent 
\begin{figure}[H]
\noindent \centering{}\includegraphics[scale=0.4]{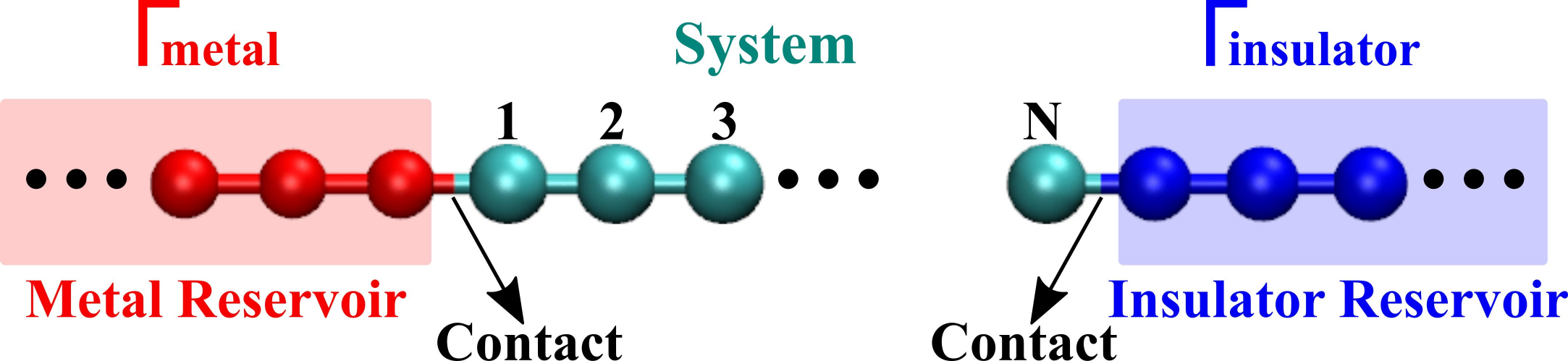}\caption{\label{fig:linearchain}Schematic representation of the one-dimensional model we use to describe a domain wall.
The central $N$-site sector, denoted as $H_{S}$ in the Hamiltonian,
contains the domain wall. It is attached to semi-infinite leads on
both sides; a uniform, strongly correlated metal on the left and a
uniform Mott insulator on the right. Both these reservoirs are described
by the $H_{R}$ term in the Hamiltonian. The $N$-site chain system
is connected to the reservoirs through contact components, denoted
as $H_{C}$ in the Hamiltonian.}
\end{figure}

To make our notation transparent, it is convenient to separate the Hamiltonian in three terms (see
Fig.~\ref{fig:linearchain})
\begin{equation}
H=H_{S}+H_{R}+H_{C}.\label{eq:hamiltonian2}
\end{equation}
The first term $H_{S}$ is a Hubbard Hamiltonian for the $N$ central
sites (``system'')
\begin{eqnarray}
H_{S} & = & -t\sum_{i=1,\sigma}^{N-1}\left[c_{i\sigma}^{\dagger}c_{\left(i+1\right)\sigma}^{\phantom{\dagger}}+\mathrm{h.c.}\right]\nonumber \\
 &  & +U\sum_{i=1}^{N}\left(n_{i\uparrow}-\frac{1}{2}\right)\left(n_{i\downarrow}-\frac{1}{2}\right),\label{eq:hs}
\end{eqnarray}
$H_{R}$ refers to the semi-infinite chains to the left and to the
right of the system (``reservoirs'')
\begin{eqnarray}
H_{R} & = & -t\left(\sum_{i=-\infty,\sigma}^{-1}+\sum_{i=N+1,\sigma}^{\infty}\right)\left[c_{i\sigma}^{\dagger}c_{\left(i+1\right)\sigma}^{\phantom{\dagger}}+\mathrm{h.c.}\right]\nonumber \\
 &  & +U\left(\sum_{i=-\infty}^{0}+\sum_{i=N+1}^{\infty}\right)\left(n_{i\uparrow}-\frac{1}{2}\right)\left(n_{i\downarrow}-\frac{1}{2}\right),\label{eq:hr}
\end{eqnarray}
and $H_{C}$ represent the coupling (``contacts'') of the central system to the reservoirs 
\begin{equation}
H_{C}=-t\sum_{\sigma}\left[c_{0\sigma}^{\dagger}c_{1\sigma}^{\phantom{\dagger}}+c_{N\sigma}^{\dagger}c_{\left(N+1\right)\sigma}^{\phantom{\dagger}}+\mathrm{h.c.}\right].\label{eq:hc}
\end{equation}

\begin{figure}[H]
\centering{}\includegraphics[scale=0.3]{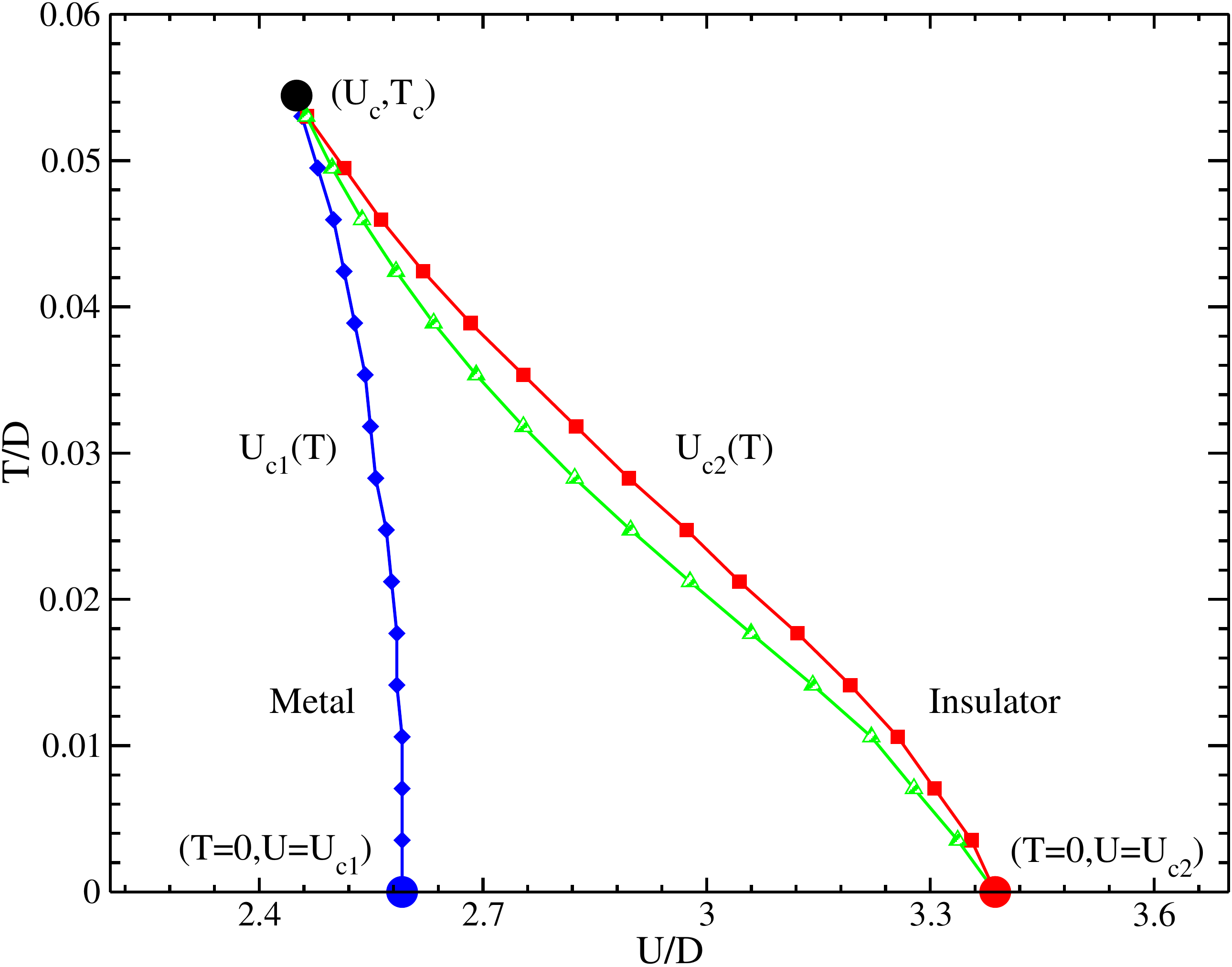}\caption{\label{fig:spinodal}DMFT phase diagram for the Mott transition at half-filling, obtained using IPT as the impurity solver. The phase coexistence region is bounded by the spinodal lines $U_{c1}(T)$ (blue line) and
$U_{c2}(T)$ (red line) where the respective insulating and metallic solutions become
unstable.  The green line marks the first order transition line $U_{c}(T)$, where the free energies of the two phases coincide.}
\end{figure}

In the following, we solve this model using Dynamical Mean-Field Theory (DMFT).\cite{MetznerVollhardt89,A.Georges1992a,Pruschke1995,A.Georges1996}
The essential simplification of this approach is the assumption that
the single-particle self-energy is a local but frequency-dependent quantity.\cite{MetznerVollhardt89}
This self-energy is then calculated from the solution of an ensemble of auxiliary single-impurity
problems supplemented with appropriate self-consistency conditions.\cite{A.Georges1992a}
Within DMFT, the Mott transition at half-filling exhibits a
coexistence region where both the metal and the insulator represent
locally stable thermodynamic phases. This coexistence region is delimited
in the $T$ vs. $U$ phase diagram by two spinodal lines, $U_{c1}(T)$ (where the insulator becomes unstable)
and $U_{c2}(T)$ (for the instability of the metal), as shown in Fig.~\ref{fig:spinodal}. 
We further concentrate on behavior along the first order transition line $U_{c}(T)$ (green line in Fig. \ref{fig:spinodal}) where the
free energies of the respective phases become equal. \cite{H.Terletska,moeller}
To describe domain wall formation,\cite{Nigel} we impose boundary conditions such that the
sites to the left of the system are in the metallic
phase, whereas sites to the right of it correspond to the Mott insulator. 
The intermediate region will then have to smoothly interpolate between
metal and insulator, thus producing a domain wall between the two
phases. 

Naturally, we will no longer be able to assume a uniform, i.e., site-independent
self-energy throughout the system. We will thus generalize the assumptions
of DMFT to accommodate a non-uniform albeit still site-diagonal self-energy
function
\begin{equation}
\Sigma\left(\omega\right)\to\Sigma_{i}\left(\omega\right).\label{eq:selfenergy}
\end{equation}
This approach was first proposed in the context of a disordered system
in Refs. \onlinecite{Dobrosavljevi'c1997,Dobrosavljevi'c1998}
(for a review, see Ref. \onlinecite{2012book}). In the following, we explain the details of
how the self-energy is computed for the present model. Like in the homogeneous DMFT, we
focus on site $i$, whose dynamics, we assume, is that of a single
correlated impurity site embedded in a bath of conduction electrons,
whose action in imaginary time is
\begin{eqnarray}
S_{eff}\left(i\right) & = & \sum_{\sigma}\int_{0}^{\beta}d\tau c_{i\sigma}^{\dagger}\left(\tau\right)\left(\partial_{\tau}-U/2\right)c_{i\sigma}\left(\tau\right)\nonumber \\
 & + & \sum_{\sigma}\int_{0}^{\beta}d\tau\int_{0}^{\beta}d\tau^{\prime}c_{i\sigma}^{\dagger}\left(\tau\right)\Delta_{i}\left(\tau-\tau^{\prime}\right)c_{i\sigma}\left(\tau^{\prime}\right)\nonumber \\
 & + & U\int_{0}^{\beta}d\tau n_{i\uparrow}\left(\tau\right)n_{i\downarrow}\left(\tau\right).\label{eq:effaction}
\end{eqnarray}
The hybridization function $\Delta_{i}\left(\tau-\tau^{\prime}\right)$
describes processes whereby an electron hops out of site $i$ at time
$\tau^{\prime}$, wanders through the rest of the lattice, and hops
back onto $i$ at a later time $\tau$. We will specify how it is
computed shortly. The local Green's function of the impurity described
by the action of Eq.~(\ref{eq:effaction}) is 
\begin{equation}
G_{i}\left(\tau\right)=-\left\langle T\left[c_{i\sigma}\left(\tau\right)c_{i\sigma}^{\dagger}\left(0\right)\right]\right\rangle _{eff},\label{eq:greenloc}
\end{equation}
where the subscript $eff$ emphasizes that it should be calculated
under the dynamics of Eq.~(\ref{eq:effaction}). The self-energy
$\Sigma_{i}\left(i\omega_{n}\right)$ is then obtained from the Fourier
transform to Matsubara frequencies

\begin{equation}
G_{i}\left(i\omega_{n}\right)=\frac{1}{i\omega_{n}+U/2-\Delta_{i}\left(i\omega_{n}\right)-\Sigma_{i}\left(i\omega_{n}\right)}.\label{eq:selfen}
\end{equation}
The lattice single-particle Green's function is given within this
scheme by the resolvent (we use a hat to denote a matrix in the lattice
site basis)
\begin{equation}
\widehat{G}\left(i\omega_{n}\right)=\left[i\omega_{n}-\widehat{H}_{0}-\widehat{\Sigma}\left(i\omega_{n}\right)\right]^{-1},\label{eq:resolvent}
\end{equation}
where $\widehat{H}_{0}$ is the non-interacting Hamiltonian {[}Eq.~(\ref{eq:hamiltonian})
with $U=0${]} and the matrix elements of the self-energy operator
$\widehat{\Sigma}\left(i\omega_{n}\right)$ in the site basis is
\begin{equation}
\left\langle i\left|\widehat{\Sigma}\left(i\omega_{n}\right)\right|j\right\rangle =\Sigma_{i}\left(i\omega_{n}\right)\delta_{ij}.\label{eq:selfenoperator}
\end{equation}
The self-consistency loop is closed by requiring that the site-diagonal
elements of the lattice Green's function coincide with the local Green's
functions of Eq.~(\ref{eq:greenloc})
\begin{equation}
\left\langle i\left|\widehat{G}\left(i\omega_{n}\right)\right|i\right\rangle =\frac{1}{i\omega_{n}+U/2-\Delta_{i}\left(i\omega_{n}\right)-\Sigma_{i}\left(i\omega_{n}\right)}.\label{eq:selfconsstatDMFT}
\end{equation}
This last equation provides an expression for a self-consistent hybridization
function for each site $\Delta_{i}\left(i\omega_{n}\right)$. In a
completely homogeneous situation, the above scheme reduces to the standard
DMFT.

The procedure described above requires the computation of the local
Green's function of Eq.~(\ref{eq:greenloc}) or, equivalently, the
self-energy defined in Eq.~(\ref{eq:selfen}) for the problem defined
by the single-impurity action of Eq.~(\ref{eq:effaction}). To this end, we  
utilized Iterated Perturbation Theory (IPT)\cite{A.Georges1992a,H.Kajueter1996}
as the required impurity solver. This procedure, while being computationally much 
more efficient than standard QMC methods, has previously been shown to properly capture both the
insulating and the metallic solutions to the problem,  as
well as most other qualitative features of the full DMFT solution,\cite{X.Y.Zhang1993}
which will suffice for our purposes. 

If the system size $N$ is large enough to accommodate the full extension
of the domain wall, the self-energy will be practically uniform in
the region of the reservoirs. In carrying out our computations for different 
values of $U$ and $T$, we  carefully verified that this condition is satisfied. 
For the parameter range explored in this work,
$N=50$ proved sufficient, except for the largest domain wall size we
analyzed, for which a value of $N=70$ was required. 

Although the system studied is infinite, the computation of the local
Green's function and self-energy in the domain wall region is all
that is required, as we now explain. It is easy to see that the non-interacting
Hamiltonian $\widehat{H}_{0}$ of the full infinite system has an
obvious block structure given by

\textcolor{black}{
\begin{equation}
\widehat{H}_{0}=\left[\begin{array}{cc}
\widehat{H}_{S} & \widehat{H}_{C}\\
\widehat{H}_{C} & \widehat{H}_{R}
\end{array}\right].\label{eq:blockham}
\end{equation}
Likewise, the self-energy can also be separated into system {[}$\widehat{\Sigma}_{S}\left(i\omega_{n}\right)${]}
and reservoir {[}$\widehat{\Sigma}_{R}\left(i\omega_{n}\right)${]}
blocks
\begin{equation}
\widehat{\Sigma}\left(i\omega_{n}\right)=\left[\begin{array}{cc}
\widehat{\Sigma}_{S}\left(i\omega_{n}\right) & 0\\
0 & \widehat{\Sigma}_{R}\left(i\omega_{n}\right)
\end{array}\right].\label{eq:blockselfen}
\end{equation}
The lattice Green's function (\ref{eq:resolvent}) satisfies the equation
\begin{equation}
\left[i\omega_{n}-\widehat{H}_{0}-\widehat{\Sigma}\left(i\omega_{n}\right)\right]\widehat{G}\left(i\omega_{n}\right)=\widehat{1},\label{eq:resolv2}
\end{equation}
where $\widehat{1}$ is the unit matrix. In block form, Eq.~(\ref{eq:resolv2})
reads}\begin{widetext}\textcolor{black}{{} 
\begin{equation}
\left[\begin{array}{cc}
i\omega_{n}-\widehat{H}_{S}-\widehat{\Sigma}_{S}\left(i\omega_{n}\right) & \widehat{H}_{C}\\
\widehat{H}_{C} & i\omega_{n}-\widehat{H}_{R}-\widehat{\Sigma}_{R}\left(i\omega_{n}\right)
\end{array}\right]\left[\begin{array}{cc}
\widehat{G}_{S} & \widehat{G}_{C}\\
\widehat{G}_{C} & \widehat{G}_{R}
\end{array}\right]=\left[\begin{array}{cc}
\widehat{1} & 0\\
0 & \widehat{1}
\end{array}\right],\label{eq:resolvmatrix}
\end{equation}
}\end{widetext}\textcolor{black}{which can be written out as
\begin{eqnarray}
\left[i\omega_{n}-\widehat{H}_{S}-\widehat{\Sigma}_{S}\left(i\omega_{n}\right)\right]\widehat{G}_{S}+\widehat{H}_{C}\widehat{G}_{C} & = & \widehat{1},\label{eq:resolvcol1}\\
\widehat{H}_{C}\widehat{G}_{S}+\left[i\omega_{n}-\widehat{H}_{R}-\widehat{\Sigma}_{R}\left(i\omega_{n}\right)\right]\widehat{G}_{C} & = & 0.\label{eq:resolvcol2}
\end{eqnarray}
}Eq.~(\ref{eq:resolvcol2}) can be solved to give
\begin{equation}
\widehat{G}_{C}=-\left[i\omega_{n}-\widehat{H}_{R}-\widehat{\Sigma}_{R}\left(i\omega_{n}\right)\right]^{-1}\widehat{H}_{C}\widehat{G}_{S},\label{eq:gc}
\end{equation}
and the result can be plugged into Eq.~(\ref{eq:resolvcol1}) to
yield

\begin{eqnarray}
\widehat{G}_{S} & = & \frac{1}{i\omega_{n}-\widehat{H}_{S}-\widehat{\Sigma}_{S}\left(i\omega_{n}\right)-\widehat{H}_{C}\left[\frac{1}{i\omega_{n}-\widehat{H}_{R}-\widehat{\Sigma}_{R}\left(i\omega_{n}\right)}\right]\widehat{H}_{C}}\\
 & \equiv & \frac{1}{i\omega_{n}-\widehat{H}_{S}-\widehat{\Sigma}_{S}\left(i\omega_{n}\right)-\widehat{\Gamma}\left(i\omega_{n}\right)},\label{eq:gs}
\end{eqnarray}
where 
\begin{equation}
\widehat{\Gamma}\left(i\omega_{n}\right)=\widehat{H}_{C}\left[\frac{1}{i\omega_{n}-\widehat{H}_{R}-\widehat{\Sigma}_{R}\left(i\omega_{n}\right)}\right]\widehat{H}_{C},\label{eq:gamma}
\end{equation}
contains all the information from the reservoirs needed for the calculation
of the system's Green's function. Since the self-energy in the reservoirs
is site-independent, we can write, for the one-dimensional lattice
we are using,

\begin{equation}
\widehat{\Gamma}(i\omega_{n})=\widehat{\Gamma}_{L}(i\omega_{n})+\widehat{\Gamma}_{R}(i\omega_{n}),\label{eq:gamma1d}
\end{equation}
where $\widehat{\Gamma}_{L(R)}\left(i\omega_{n}\right)$ is the contribution
from the reservoir to the left (right) of the system. The latter quantities
are given in terms of the (purely imaginary) self-energies on the
left (metal) and right (insulator) $\Sigma_{L,R}\left(i\omega_{n}\right)$
as 
\begin{eqnarray}
\widehat{\Gamma}_{L(R)}(i\omega_{n}) & = & \frac{i}{2}\left\{ \Omega_{L(R)}(i\omega_{n})\phantom{\sqrt{\left[\Omega_{L(R)}(i\omega_{n})\right]^{2}+4t\text{\texttwosuperior}}}\right.\nonumber \\
 &  & \left.-\mathrm{sgn}\left(\omega_{n}\right)\sqrt{\left[\Omega_{L(R)}(i\omega_{n})\right]^{2}+4t\text{\texttwosuperior}}\right\} ,\label{eq:gamma3}
\end{eqnarray}
where $\Omega_{L(R)}(i\omega_{n})=\omega_{n}-\mathrm{Im}\Sigma_{L(R)}(i\omega_{n})$.

In all our calculations we analytically continue the Matsubara Green's
functions and self-energies to the real axis using the Pad\'{e} approximation.\cite{serene}

\section{\textit{\emph{Results}}}

\label{sec:results}

Next, we present the detailed results we obtained, exploring the behavior of a domain wall within the 
coexistence region, in the entire range of temperatures
$0 <T < T_c$, along the first-order transition line. 

\begin{figure}[H]
\includegraphics[scale=0.30]{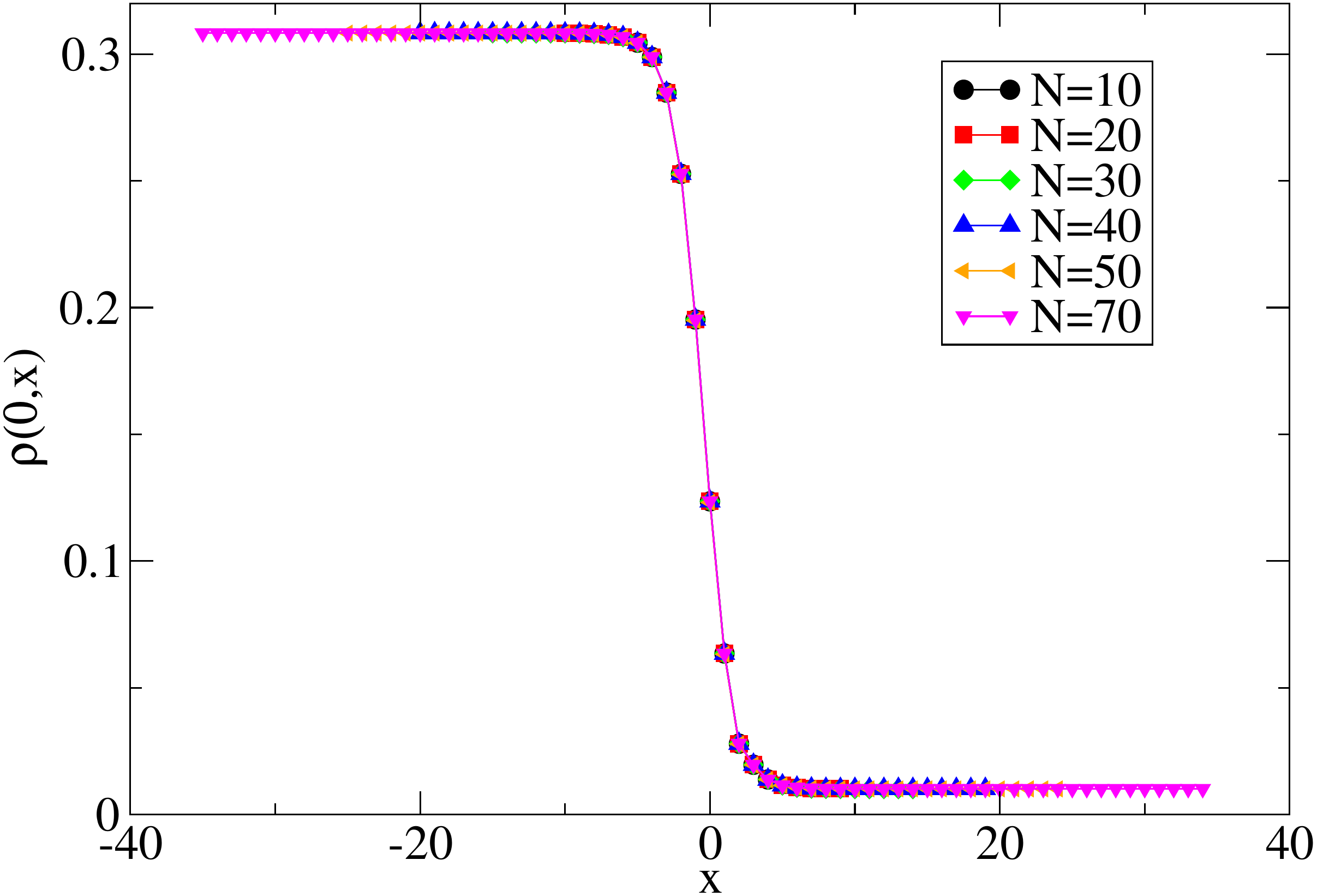}
\caption{Spatial variation of the local DOS $\rho(0,x)$ at the Fermi energy, across the domain separating a strongly correlated metal (left) and a Mott insulator (right), corresponding to $T/D=0.0351$ and $U/D=2.69773$. Results are shown for different system sizes $N$ used in our simulation, demonstrating negligible finite-size effects of our results.}
\label{fig:sizedependence}
\end{figure}

As we mentioned in Section~\ref{sec:method}, an accurate calculation needs to make sure that
the system size $N$ is large enough so that our position-dependent solution converges to the 
proper asymptotic limit (``reservoirs''), where the spatial variation can be ignored. To illustrate this,
in Fig.\ref{fig:sizedependence} we display the domain wall profile, 
as described by the spatial variation of the local density of states (DOS) $\rho(0,x)= -(1/\pi ) \mathrm{Im}G(\omega=0,x)$, evaluated at $T/D=0.035$
and $U/D=2.697$, and plotted as a function of the coordinate $x$ perpendicular to the domain wall. 
This quantity is small in the insulator (approaching zero as $T\to0$), but remains finite in the metal, thus displaying strong spatial variation across the domain wall. As we can see in
\noindent  Fig.\ref{fig:sizedependence},
 the spatial profile of the domain wall displays very little change with the size of the central region ($N$ sites), where we allow for spatial variation. This means that our system size  $N$ 

\begin{figure}[H]
\begin{centering}
\includegraphics[scale=0.3]{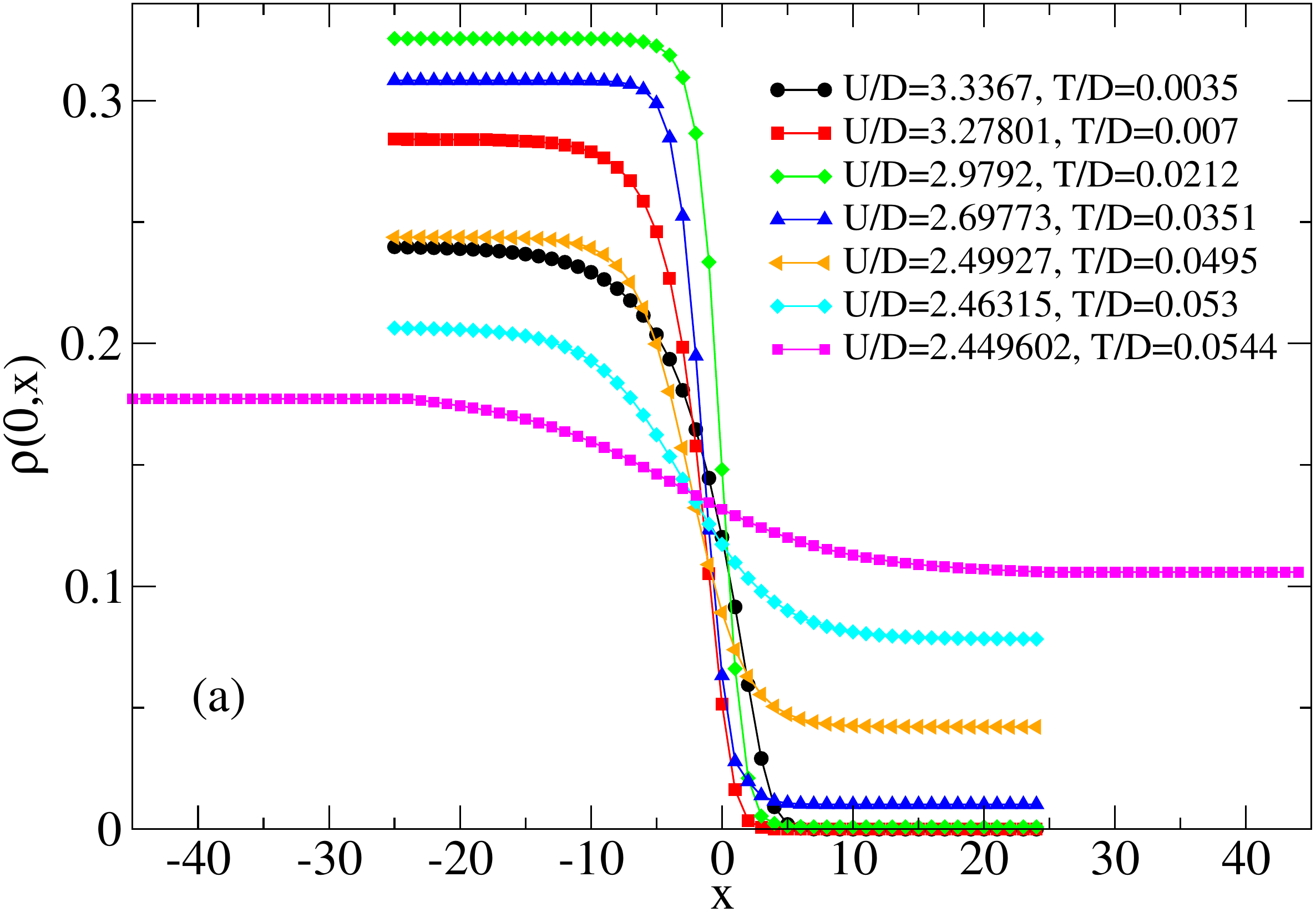}\\
\includegraphics[scale=0.3]{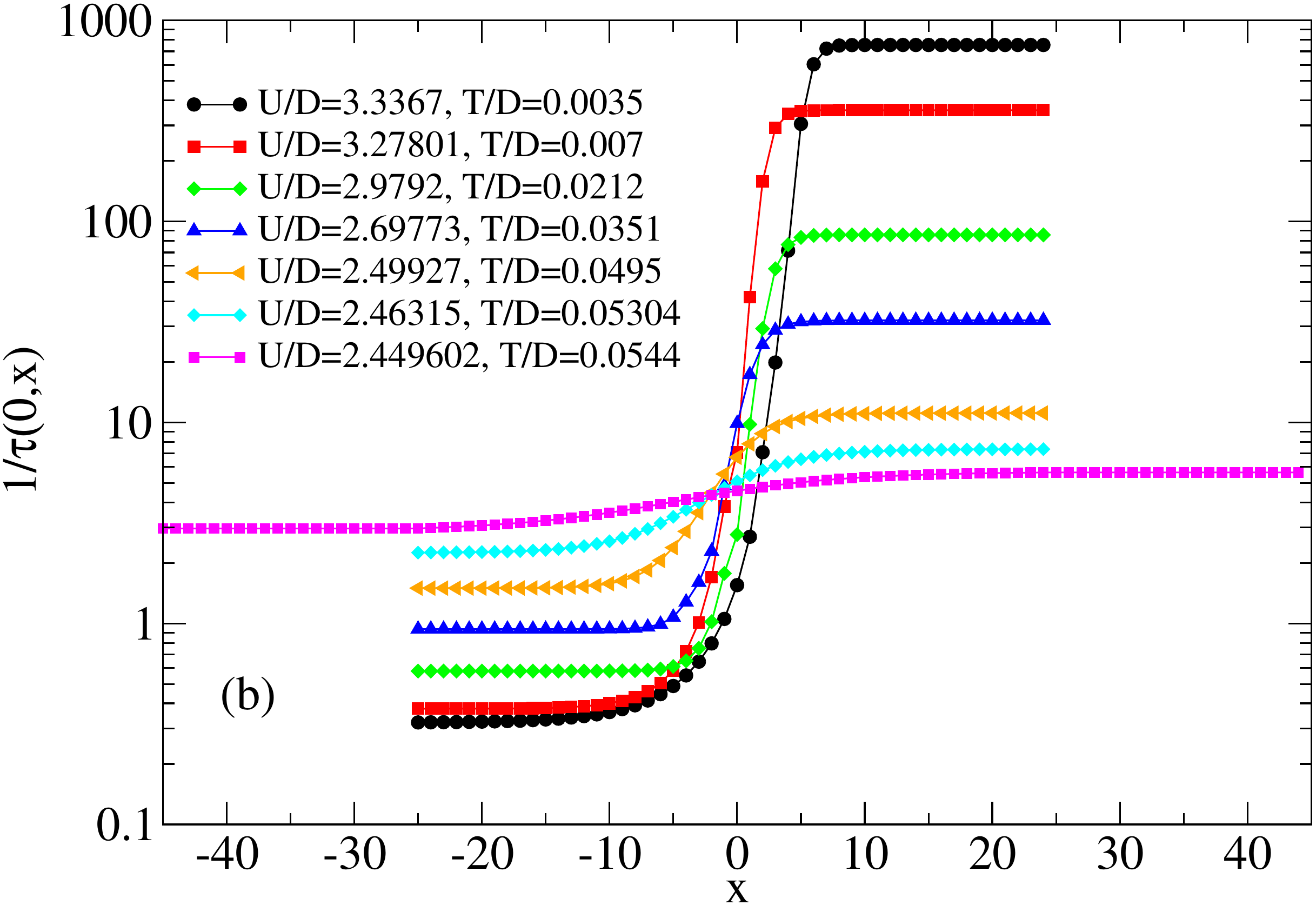}

\par\end{centering}
\caption{ Variation of the domain wall profile with temperature:  The local DOS (a) and the  scattering rate (b) shown for different $T$ and $U$, following the first-order transition line. Note a slightly non-monotonic dependence on temperature.}
\label{fig:dw_a}
\end{figure}

\begin{figure}[H]
\begin{centering}
\includegraphics[scale=0.3]{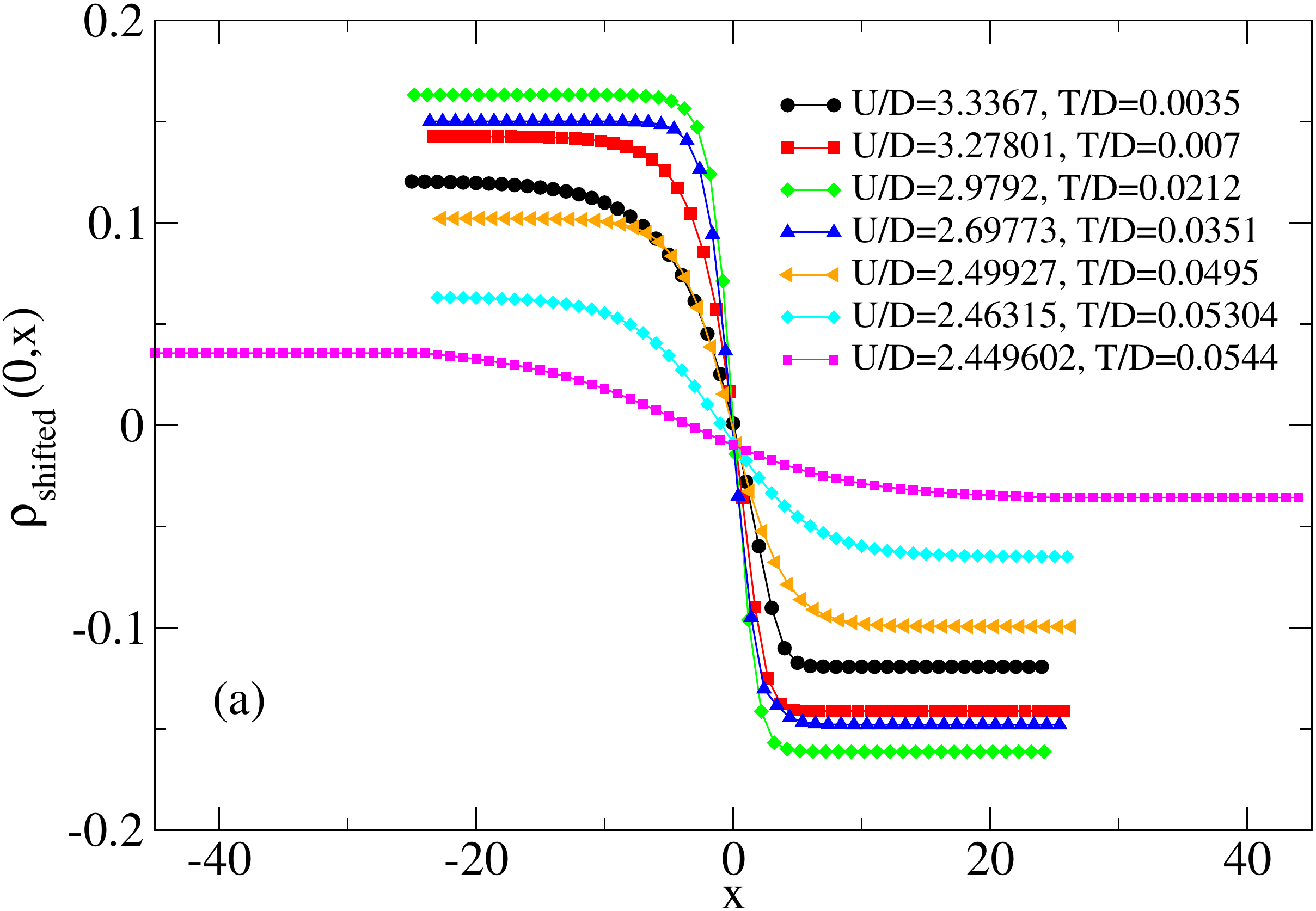}\\
\includegraphics[scale=0.3]{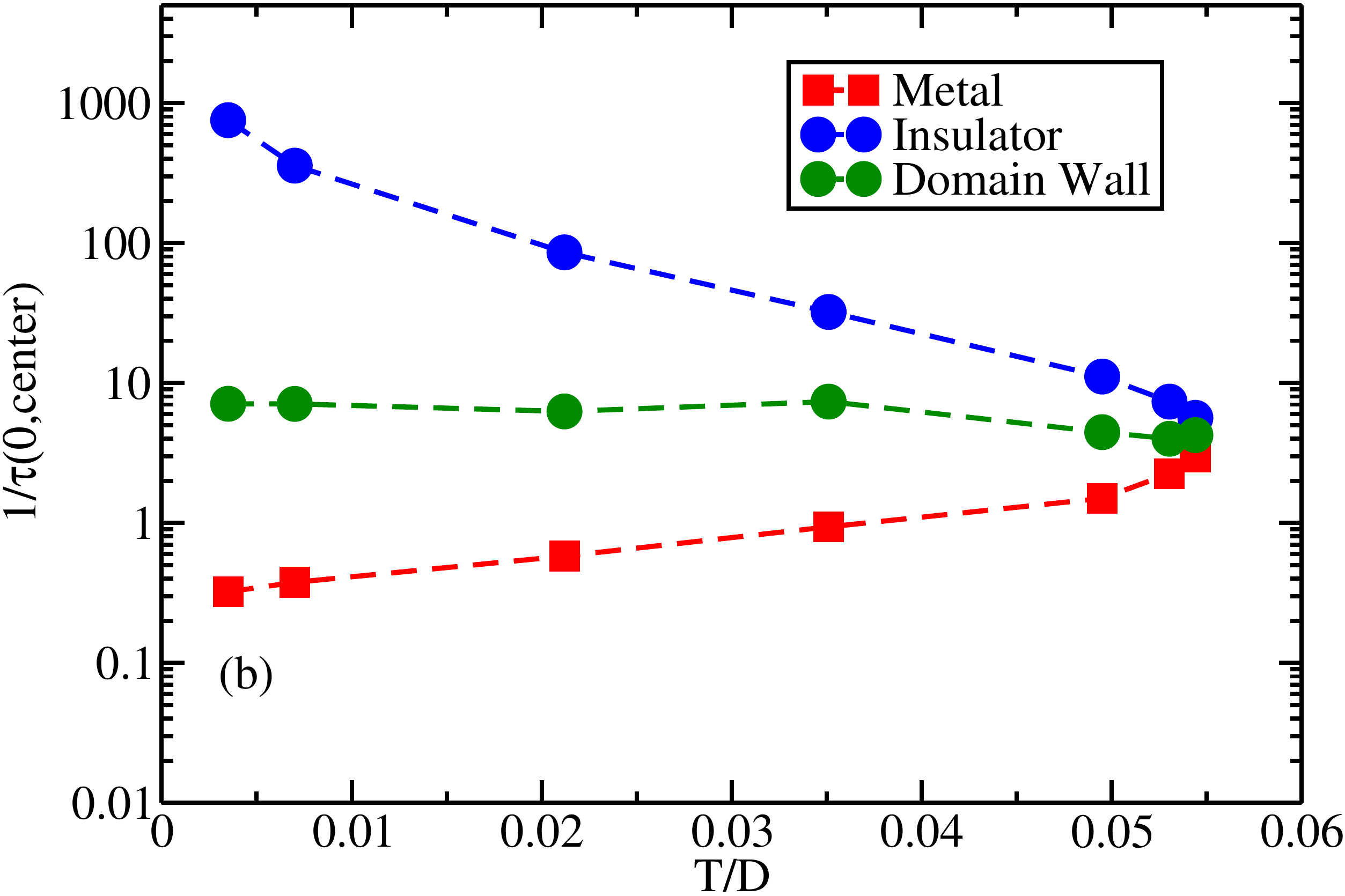}\\
\par\end{centering}
\caption{(a) Variation of the domain wall profile with temperature. Same data are shown in
Fig.~\ref{fig:dw_a}, but here they have been shifted both horizontally
and vertically so that the domain wall center is located at the plot
center. This way of plotting reveals more clearly the variation of the {\em domain wall thickness} as a function of temperature. (b) The scattering rate at the domain wall center ($x=0)$ displays very weak temperature dependence (central curve), in dramatic contrast to the behavior of either the insulator (top curve), or the metal (bottom curve) evaluated for same $T$ and $U$, along the first-order line.
}
\label{fig:dw-shifted}
\end{figure}
\noindent is large enough to eliminate any finite-size effects from our calculation. Performing similar calculations for the  different temperatures, we verified
that $N=70$ is sufficient for an accurate description at all the relevant temperatures
   ($0 < T< T_{c}$) within the coexistence region. 
   
\subsection{Anomalous dynamics of the domain walls}
   
For each temperature we considered, we selected the precise value of $U(T)$ that falls on the 
first order transition line (see green line of Fig.~\ref{fig:spinodal}). Results obtained for several temperatures are shown in Fig.~\ref{fig:dw_a} (a), showing the domain-wall profiles of  the local DOS. We should mention that, 
within our simulation, the precise position of the domain wall we find for given $T$ is somewhat sensitive to the exact value for $U(T)$ selected. We had to, accordingly, adjust the value of $U$ to a  precision of several 
decimal places, in order to obtain adequate center alignment, which is helpful for comparing the detailed form of the domain wall profile at different temperatures. For better comparison, in Fig.~\ref{fig:dw-shifted} (a)
we display the same data translated along both the $x$- and the $y$-axes, so that the domain wall center coincides with the coordinate origin. Note that the local DOS in the uniform regions has a non-monotonic behavior with $T$, a behavior we expand upon in the Appendix.

\begin{figure}[H]
\includegraphics[scale=0.3]{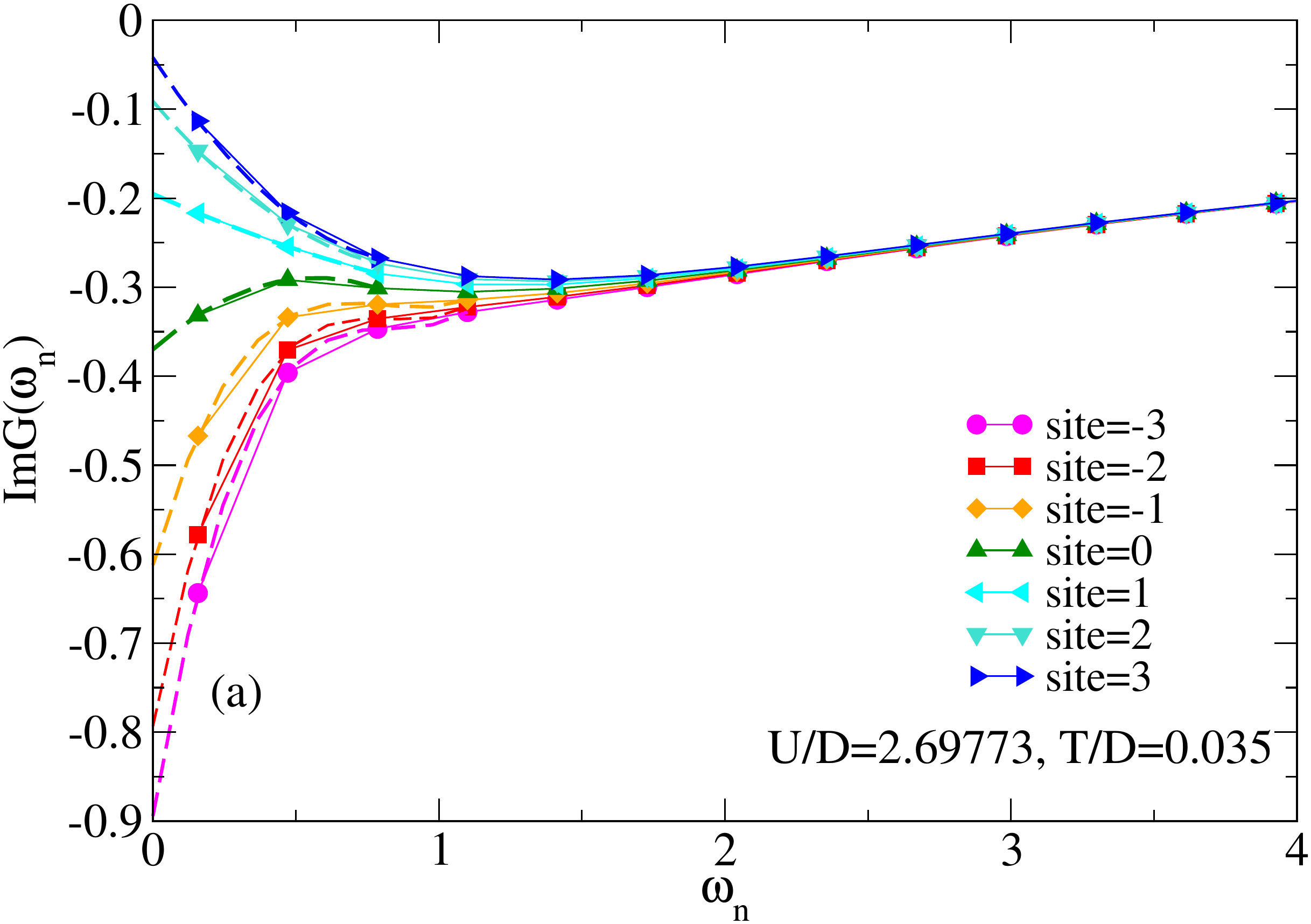}\\
\includegraphics[scale=0.3]{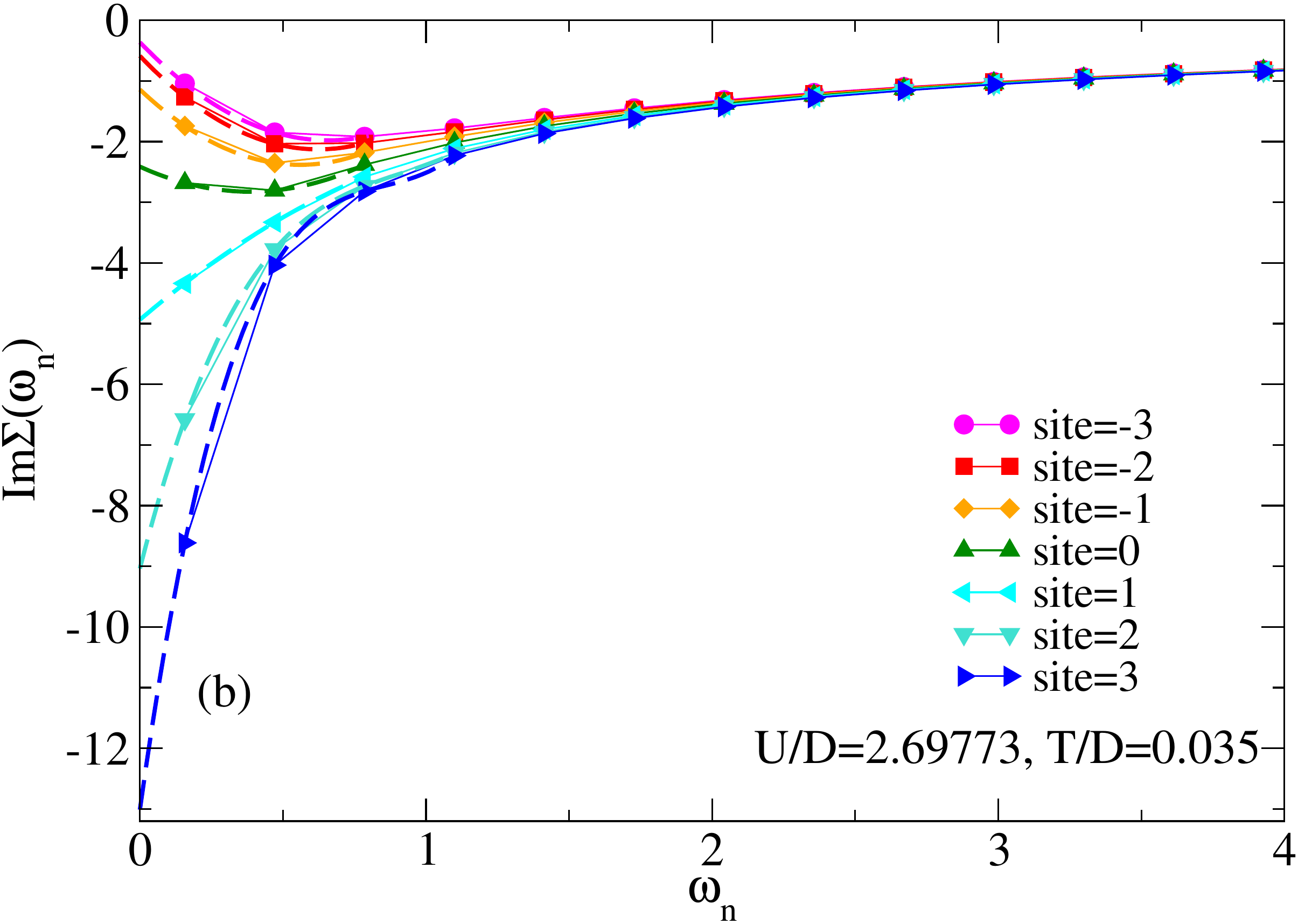}
\caption{\label{fig:dw_freqdep}Domain wall central region: \textcolor{black}{Matsubara
frequency dependence of the imaginary part of (a) the local Green's
function and (b) the self-energy for some sites in the central portion
of the domain wall at $T/D=0.035$} and $U/D=2.69773$.}
\end{figure}

The corresponding behavior of the inelastic scattering rate 
$1/\tau (0,x) = -2 \mathrm{Im}\; \Sigma(\omega = 0,x)$ across the domain wall and different temperatures is shown in Fig.~\ref{fig:dw_a} (b). It is generally expected to be small in a coherent metal 
(in a Fermi liquid $1/\tau \sim T^2$ for given $U$)  and very large in a Mott insulator, as we 
observe in the respective phases. This behavior reflects a fundamentally different nature 
of transport in the two competing phases, but even more interesting behavior is seen within the domain wall
itself. Here the scattering rate $1/\tau$ smoothly interpolates between the two limits and thus retains very weak T-dependence, reflecting significant electron-electron scattering down to the lowest temperatures! This surprising result is displayed even more precisely by plotting $1/\tau$ evaluated at the domain wall center as a function of temperature, in comparison to the behavior of the two phases, as shown in Fig.~\ref{fig:dw-shifted} (b). 
Similar behavior is also seen in the frequency dependence of the corresponding 
Green's function and the self-energy, shown as a function of the Matsubara frequency in
Fig.~\ref{fig:dw_freqdep}, for several sites across  the domain wall
at $T=0.035D$ and $U=2.698D$. Here we observe a characteristic evolution  
from metallic to insulating behavior, as one moves across
the domain wall, which is most pronounced at the lowest 
frequencies. For the sites at the center of the domain wall, however, we
observe characteristically weak frequency dependence. This behavior is clearly distinct from either a metal or an insulator, but is constrained by having to interpolate from one to the other.

The domain wall center is, therefore, recognized as an {\em incoherent conductor} down to the lowest temperatures. Physically, such  {\em non-Fermi liquid} behavior makes it clear that the domain wall represents a {\em different state of matter} from either a coherent (Fermi liquid) metal, or a Mott insulator. This surprising result could be regarded as a curiosity with little physical consequence in situations where the relative volume (area) fraction ``covered'' by domain walls is negligibly small compared to the bulk of the system. In the presence of sufficient disorder, however, both recent simulations\cite{martha} and experiments\cite{M.Qazilbash} demonstrate a surprisingly abundant proliferation of such domain walls, suggesting a fundamentally new physical picture. We may expect this to be especially significant whenever the domain walls themselves are sufficiently fat (thick), so that a sizeable fraction of the system's volume (area) is affected by such ``resilient'' inelastic electron-electron scattering, which persists to low temperatures, in contrast to the behavior expected for conventional metals.



\subsection{What controls the thickness of the domain walls?}
 
To precisely quantify the domain wall thickness as a function of temperature, we fit its shape 
to the standard $\tanh(x/\xi)$ form, generally found for domain
walls separating two coexisting phases \cite{Nigel}. To be more precise, such
symmetric domain walls of thickness given by an appropriate correlation length
$\xi$ is what one expects near any finite-temperature critical end-point at $T=T_c$, as we also find. At lower temperatures, however, our two phases 
are not related by any static symmetry, hence the domain wall should not necessarily retain its symmetric form, since the correlation length of the  respective phases may not be exactly the same. Indeed, even a quick look at Fig.~\ref{fig:dw-shifted} (a) reveals that at lower temperatures, the domain walls are 
much ``thicker'' on the metallic than on the insulating side. 
 
\begin{figure}[H]
\begin{centering}
\includegraphics[scale=0.23]{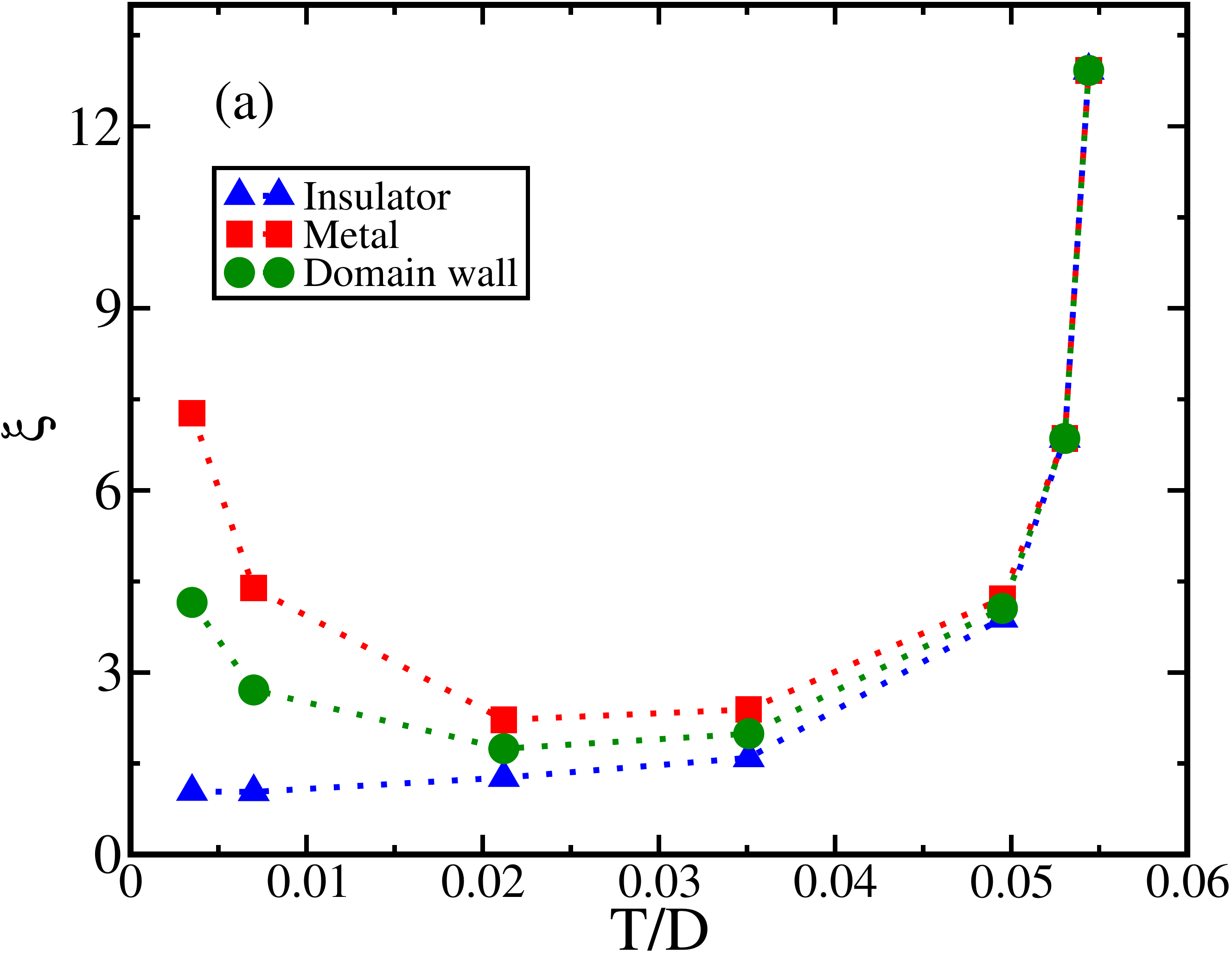}\\
\includegraphics[scale=0.23]{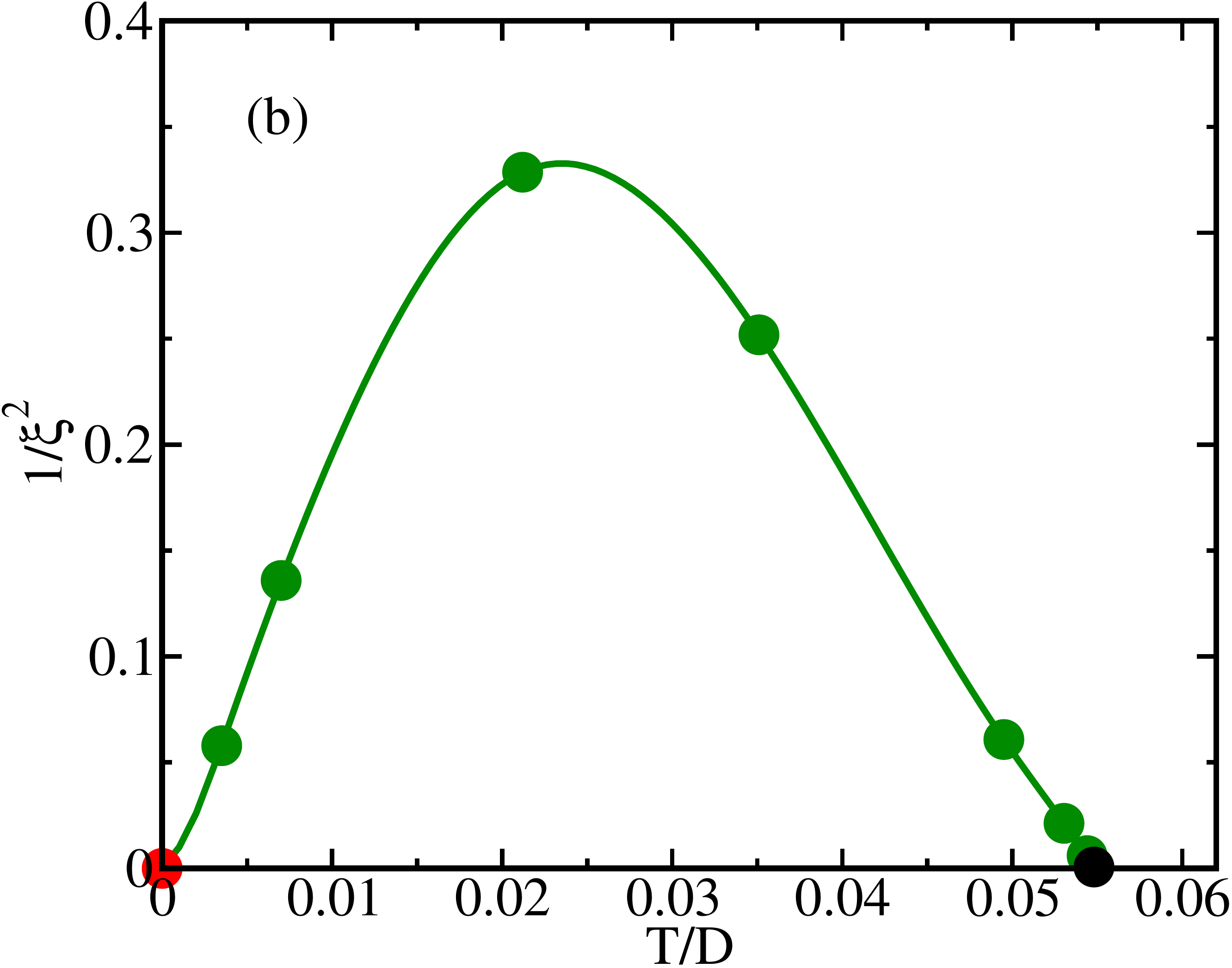}
\par\end{centering}
\caption{(a) Domain wall thickness as a function of temperature, displaying pronounced asymmetry at low temperature, where the thickness on the metallic side $\xi_{met}$ (red squares) becomes much larger than its insulating counterpart $\xi_{ins}$ (blue triangles). (b) The overall thickness $\xi = (\xi_{met} + \xi_{ins})/2$ (green circles in both panels) displays a square-root singularity at both $T=T_c$ (black dot) and the $T=0$ critical point (red dot). }
\label{fig:dw-width}
\end{figure}

To quantify 
this behavior, we perform partial fits to the $\tanh(x/\xi_a)$ form on each 
side of the domain wall center, which we define as the corresponding inflection point in its
profile. Here $\xi_a$, with $a=$ $met$ or $ins$ defines the two different correlation
lengths, corresponding to the respective metallic or insulating phase. 
The resulting $T$-dependence of $\xi_{met}$ and $\xi_{ins}$ is shown in 
Fig.~\ref{fig:dw-width} (a), together with the total domain wall thickness 
$\xi = (\xi_{met} + \xi_{ins})/2$. General arguments\cite{Nigel} predict this quantity to 
diverge at $T \rightarrow T_c$, as we find. Indeed, the critical
point at $T=T_{c}$ is known to belong to the Ising universality class
\cite{PhysRevLett.43.1957,Rozenberg1999a,Kotliar2000,Limelette03,PhysRevLett.100.026408,PhysRevLett.114.106401}.
According to an appropriate Landau theory\cite{Rozenberg1999a} for this critical point, the domain wall
width should be proportional to the corresponding correlation length, diverging at the
critical point as
$\xi\sim\xi_{corr}\sim\left|T-T_{c}\right|^{1/2}.\label{eq:dw-widthLandau}$

Remarkably, however, we find $\xi$ to display a divergence also at $T \rightarrow 0$, 
thus retaining a sizeable thickness even at intermediate temperatures. This behavior
is seen even more precisely by plotting $\xi^{-2}$ as a function of $T$ in
Fig.~\ref{fig:dw-width} (b), displaying the expected square-root divergence \cite{Nigel}
not only at at $T=T_c$ but also at $T=0$. From the practical point of view, this curious result is important, because it suggest that domain  walls should retain substantial thickness throughout the coexistence region, therefore introducing a potentially significant new feature of transport properties near the Mott point. 

What could be the mechanism leading to this strange behavior? An important clue is provided by comparing the behavior of the corresponding correlation length describing the domain wall profile, as shown in Fig.~\ref{fig:dw-width} (a). Here we observe that, while both $\xi_{met}$ and $\xi_{ins}$ diverge (and coincide)  at $T \rightarrow T_c$, they behave very differently at $T \rightarrow 0$. Here  $\xi_{met}$ diverges, but 
$\xi_{ins}$ saturates to a small value comparable to one lattice spacing. Physically, this result is easy to understand, keeping in mind the nature of the critical point at $U = U_{c2}$, which we approach as $T\rightarrow 0$ along the first order line. This critical point signals the {\em instability of the metallic phase}, where the characteristic energy scale of the quasi-particles vanishes and the free energy minimum corresponding to the metallic phase becomes unstable, leading to the divergence of $\xi_{met}$. In contrast, the insulating solution here remains stable, as its own instability arises only at a much smaller $U= U_{c1} \ll U_{c2}$, and the corresponding $\xi_{ins}$ thus remains short, as we find.

The resulting behavior of the overall domain wall thickness $\xi = (\xi_{met} + \xi_{ins})/2$ is even more clearly seen by plotting $\xi^{-2}$ as a function of temperature, which is seen to linearly vanish both at $ T = T_c$ and $T=0$, as shown in 
Fig.~\ref{fig:dw-width} (b). While domain walls are generally expected\cite{Nigel} to become thick at finite-temperature critical end-points ($T=T_c$), the presence of such behavior also at low temperatures deserves further comment and a proper physical interpretation. Within our DMFT formulation, it reflects the emergence of an additional critical point at $T=0$ and $U=U_{c2})$, corresponding to the divergence of the quasi-particle effective mass $m^* \sim (U_{c2} - U)^{-1}$, signaling a singular enhancement of the Sommerfeld specific-heat coefficient $\gamma = C/T \sim m^*$. This result, which is well-established within DMFT, reflects the approach to the Mott insulator characterized by large spin entropy at low temperatures. Physically, such neglect of significant inter-site spin correlations, as implied by the DMFT approximation, is expected to be justified in the limit of strong magnetic frustration, possibly in materials with triangular or Kagom\'{e} lattices.

\section{Conclusions}

In this paper we performed a detailed study of the structure and the dynamics of domains walls expected within the phase coexistence region around the Mott point. Our results, obtained within the DMFT approximation, suggest that such domain walls should display unusual dynamics, which is unlike that of a metal or that of an insulator, locally retaining strong inelastic (electron-electron) scattering down to very low temperatures. This curious behavior could be significant in systems where weak disorder and low dimensionality conspire to produce a substantial concentration of domain walls within the metal-insulator phase coexistence region. This behavior should be especially significant in systems where the domain walls remain sufficiently thick or fat over an appreciable temperature range, such that the domain wall matter covers a substantial volume (area) of a given sample.  Our predictions could be even more directly tested by STM (scanning tunneling spectroscopy) experiments, which are able to locally probe transport properties at the center of a given domain walls, in even  simpler geometries. 

Our analysis also revealed that the mechanism favoring such thick domain walls is directly related to the degree of magnetic frustration characterizing the incipient Mott insulating state. In spatially inhomogeneous systems (e.g. due to lattice defects of other forms of structural disorder), one can imagine local regions with varying degrees of local magnetic frustration. The physical picture we put forward indicates direct consequences for the structure of the corresponding domain walls, with their local thickness being a direct measure of the local magnetic frustration. The work we presented in this paper is only the first step in the investigation of situations where the interplay of phase coexistence, strong correlations, and magnetic frustration should lead to exotic forms of dynamics of electrons, but more detailed investigations along these lines remain challenges for the future. 

\section{acknowledgments}

We thank Hanna Terleska for helpful discussions. We acknowledge support
by CNPq (Brazil) through Grants No. 307041/2017-4 and No. 590093/2011-8,
Capes (Brazil) through grant 0899/2018 (E.M.) and by the  Texas Center for Superconductivity at
the University of Houston (M.Y.S.V and J.H.M). Work in Florida (V. D. and T.-H. L.) 
was supported by the NSF Grant No. 1822258, and the National High Magnetic Field Laboratory 
through the NSF Cooperative Agreement No. 1157490 and the State of Florida.

\section*{appendix: Moving along the first order line}

\begin{figure}[H]
\includegraphics[scale=0.35]{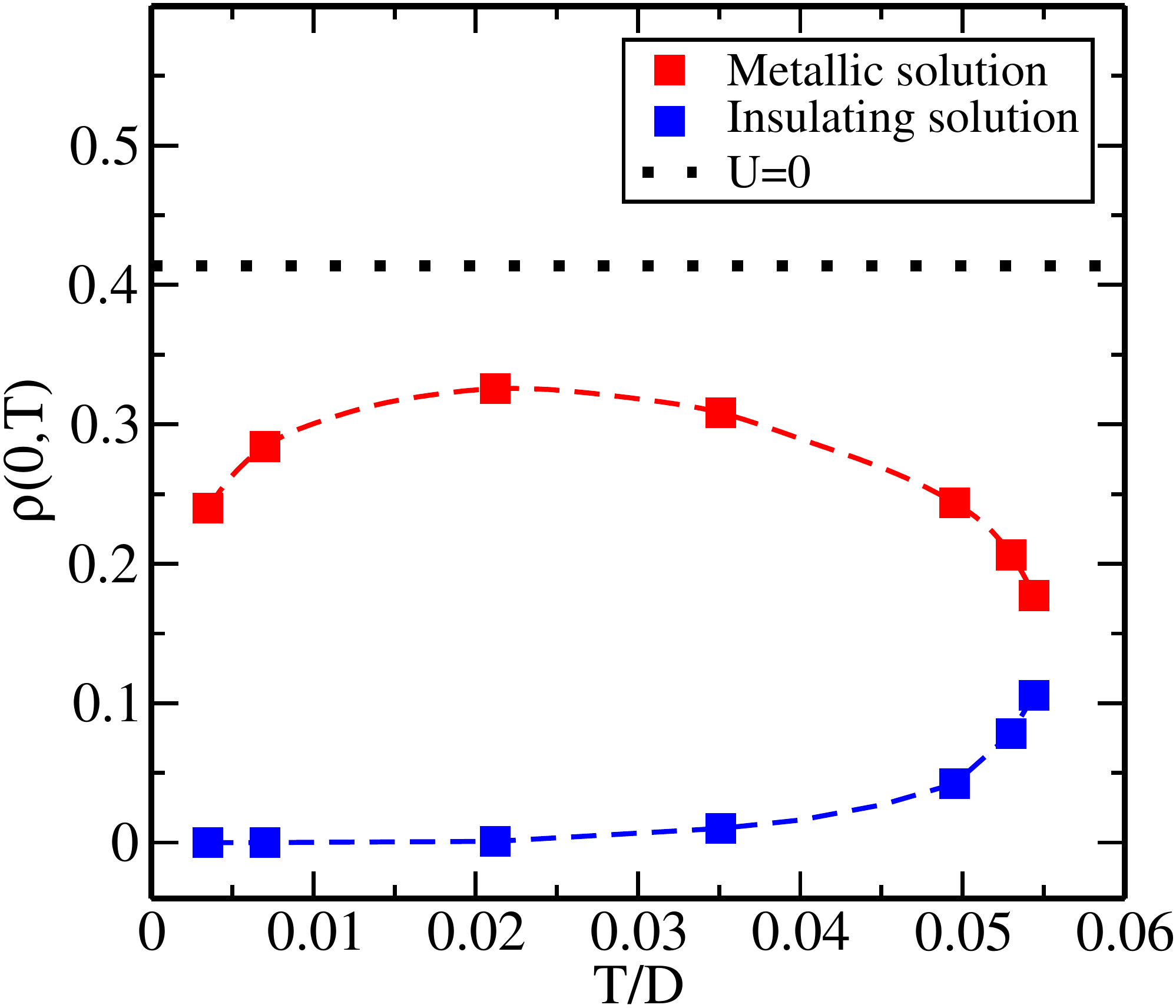}
\caption{Evolution of the density of states (DOS) along the first-order transition line, shown for both the uniform  metallic (top red curve) and the uniform insulating (bottom blue curve) solutions. For comparison we  also show the non-interacting DOS value (horizontal dotted line), which is expected for the DOS in the metallic phase strictly at $T=0$.
}
\label{fig:-green_line}
\end{figure}

A close look at the results given in Fig.~\ref{fig:dw_a} reveals some details of the  temperature dependence found, which deserve further clarification. It is clear that the DOS on the metallic side of the domain wall displays a noticeably {\em non-monotonic} $T$ dependence, which is generally not expected for a metallic phase at fixed $U$.  However, it should be noted that we have here the {\em simultaneous} variation of both $T$ and $U$ when we follow the first-order transition line (FOTL) while reducing the temperature. This complicates the analysis, producing the non-monotonic behavior, as we see even more clearly in Fig.~\ref{fig:-green_line}. Note, in particular, that the metallic DOS does not approach the non-interacting value (horizontal dotted line), even at low temperatures, in contrast to what one one finds by reducing $T$ at fixed $U$ (the so-called ``pinning condition'', not shown).

To understand this behavior, we note that, at low temperatures, the metallic phase displays Fermi liquid behavior. In this case, all quantities become scaling functions of the reduced temperature $(T/T^*_{FL})$, where  $T^*_{FL} \sim Z \sim (U_{c2} - U)$ is the Fermi liquid coherence scale and $Z$ is the quasiparticle weight.\cite{moeller} Since, within DMFT, the FOTL  {\em also} vanishes linearly  with $ (U_{c2} - U)$, the reduced temperature $(T/T^*_{FL})$ should remain finite even as $T \rightarrow 0$ along the FOTL line. This is the reason why the ``pinning condition''  is violated all along the FOTL. Indeed, within DMFT, the DOS is expected to approach its non-interacting value only at $ T \ll T^*_{FL}$, a condition that is not satisfied anywhere along the FOTL. The remaining $T$ dependence we observe represents only sub-leading corrections, which are generally complicated and non-universal, consistent with the non-monotonic behavior we find. 

\bibliographystyle{apsrev}
\bibliography{all,mit-vlad,rop}

\end{document}